\documentclass[aps,prb,twocolumn,superscriptaddress,showpacs,amsmath,amssymb]{revtex4}

\usepackage{amsfonts}
\usepackage{graphicx}
\usepackage{dcolumn}
\usepackage{bm}

\newcommand{\kagome}{kagome}
\newcommand{\nbar}{\bar{n}}

\begin{document}

\title{Exact low-temperature properties of a class of highly frustrated Hubbard models}

\author{Oleg Derzhko}
\affiliation{Institute for Condensed Matter Physics,
          National Academy of Sciences of Ukraine,
          L'viv-11, 79011, Ukraine}
\affiliation{Institut f\"ur Theoretische Physik,
          Universit\"at Magdeburg,
          P.O. Box 4120, 39016 Magdeburg, Germany}
\author{Andreas Honecker}
\affiliation{Institut f\"ur Theoretische Physik,
          Georg-August-Universit\"at G\"ottingen,
          37077 G\"ottingen, Germany}
\author{Johannes Richter}
\affiliation{Institut f\"ur Theoretische Physik,
          Universit\"at Magdeburg,
          P.O. Box 4120, 39016 Magdeburg, Germany}

\date{July 18, 2008; revised \today}


\begin{abstract}
We study the repulsive Hubbard model both analytically and numerically
on a family of highly frustrated lattices
which have one-electron states localized on isolated trapping cells.
We construct and count exact many-electron ground states for a wide range of electron densities
and obtain closed-form expressions for the low-temperature thermodynamic quantities
which are
universal
for all lattices of the family.
Furthermore, we find that saturated ferromagnetism is obtained
only for sufficiently high electron densities and large Hubbard repulsion $U$
while there is no finite average moment in the ground states at lower densities.
\end{abstract}

\maketitle

\section{Introduction}

Exactly solvable interacting quantum lattice systems are of great importance in condensed matter physics.
Although most of the examples are known for one-dimensional systems,
two-dimensional and three-dimensional models are also coming into sight.
A crucial point in searching for exactly solvable lattice models concerns the lattice geometry.
Furthermore the interplay between lattice geometry, interactions and quantum fluctuations
often gives rise to exotic quantum phases.
Famous examples where a special arrangement of interaction bonds
allows to find exact quantum many-body ground states
are the Majumdar-Ghosh model,\cite{models_mg}
the Shastry-Sutherland model,\cite{models_shasu}
and the Kitaev model.\cite{models_kitaev}

Besides revealing new magnetic properties of previously known materials
(e.g.\ the diamond-chain compound azurite \cite{azurite})
there are presently various possibilities to design interacting lattice systems with controlled geometry.
Modern strategies in chemistry open a route to synthesize new materials
with a desired lattice structure and intersite
interactions.\cite{new_materials,harrison}
Moreover, recent progress in nanotechnology allows
the fabrication of quantum dot superlattices
and quantum wire systems with any type of lattice.\cite{dots_wires}
Another rapidly developing field
is the controlled setup of optical lattices for cold atoms.\cite{optical,Wu}

Motivated by these achievements
we propose here a class of lattices
(including the well-known diamond chain,
frustrated ladder,
square-{\kagome}
and
checkerboard lattices)
for which various properties of the Hubbard model
can be examined rigorously.
In particular,
we characterize the complete manifold of highly degenerate ground states
for electron numbers $n=1,\ldots,{\cal{N}}\propto N$
and calculate low-temperature thermodynamic quantities
around a particular value of the chemical potential $\mu_0$.
The general lattice construction rules illustrated below
are based on a local point of view
similar (but not identical) to earlier considerations
for electronic \cite{flat_band_ferro,batista} and spin systems.\cite{loc_mag1}
Below we also discuss some properties of the Hubbard model 
for a few one-dimensional and two-dimensional representatives.
We illustrate the dominating role of the exact ground states
for the low-temperature physics of the Hubbard model
at certain electron densities.
Furthermore
we analyze the ground state with respect to magnetic properties.
We confirm our analytical findings by numerical data for finite systems.

\section{Lattices with trapping cells}

\label{sec:lat}

We consider the $N$-site Hubbard Hamiltonian
\begin{eqnarray}
H=\sum_{\sigma=\uparrow,\downarrow}H_{0\sigma}+H_U,
\;\;\;
H_U=U\sum_{i}n_{i,\uparrow}n_{i,\downarrow},
\nonumber\\
H_{0\sigma}=\sum_{\langle i,j \rangle}t_{i,j}
\left(c_{i,\sigma}^{\dagger}c_{j,\sigma}+c_{j,\sigma}^{\dagger}c_{i,\sigma}\right)
+\mu \sum_{i} n_{i,\sigma},
\label{01}
\end{eqnarray}
where
$i$ denotes the lattice sites,
$\langle i,j\rangle$ denote the bonds connecting neighboring sites,
the $c_{i,\sigma}^{\dagger}$ ($c_{i,\sigma}$) are the usual fermion operators,
$n_{i,\sigma} = c_{i,\sigma}^{\dagger} c_{i,\sigma}$,
$t_{i,j}>0$ are the hopping integrals,
$U\ge 0$ is the on-site Coulomb repulsion,
and $\mu$ is the chemical potential.

We consider the Hubbard model (\ref{01}) on a family of lattices
defined by the following construction rules:
(i)
Take
a `trapping cell',
i.e., a finite region  where an electron in the infinite lattice will be localized.
For simplicity,
we consider here bipartite cells with equivalent sites and bonds,
such as a single bond between two sites,
equilateral even polygons, or a cube.\cite{extension}
(ii)
Solve the one-electron problem for the trapping cell,
finding the lowest-energy eigenfunction
$\propto\sum_{i}a_{i}c_{i,\sigma}^{\dagger}\vert 0\rangle$
with $\vert 0\rangle$ denoting the vacuum state.
For traps consisting of a single bond or a square
the ground state is nondegenerate
and $a_1=-a_2=1$ or $a_1=-a_2=a_3=-a_4=1$,
respectively.
(iii)
Arrange the trapping cells into a regular pattern
so that trapping cells do not touch each other
(do not have common sites)
and complete the lattice by connecting the cells via surrounding (connecting) bonds.
Most importantly,
the connecting bond scheme should prevent the escape of the localized electron from the trap,
i.e., the constructed one-electron (localized) state
should remain an eigenstate of the Hamiltonian on the infinite lattice.
It is easy to show that a sufficient condition for this is
$\sum_{i}t_{r,i}a_i=0$,
where the sum runs over all sites $i$ of a trapping cell
and $r$ is an arbitrary site which does not belong to the trap,
see also Ref.~\onlinecite{loc_mag1}.
For such traps as a single bond, square or cube
this condition is fulfilled
if an arbitrary bond of the trapping cell
and the surrounding bonds attached to the two sites of this bond
form an isosceles triangle.
(iv)
Choose the hopping integrals $t^{\prime}$ of the trapping cells sufficiently large,
$t^{\prime} > t^{\prime}_c$,
so that the localized states become the lowest-energy ones
in comparison with other (extended) one-electron states.

Following the rules formulated above
we are able to construct many different one-dimensional, two-dimensional, 
and even three-dimensional lattices.
Note that a lattice constructed in this manner has a flat one-electron band
which for $t^{\prime} > t^{\prime}_c$ becomes the lowest-energy one.
However,
we emphasize
that such lattices do not belong to Tasaki's models for flat-band ferromagnetism.\cite{flat_band_ferro}

\begin{figure}
\begin{center}
\includegraphics[width=0.8\columnwidth]{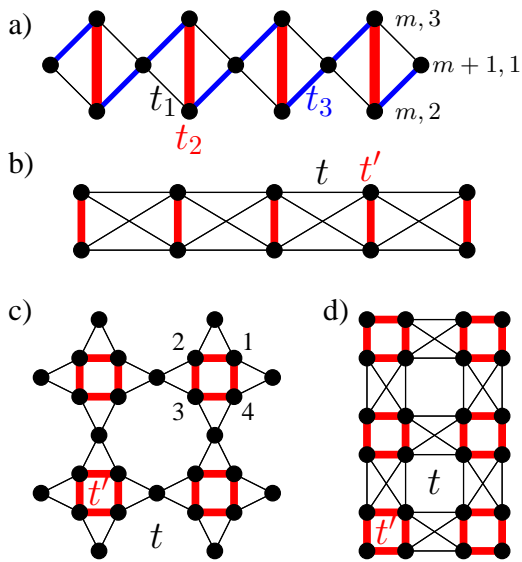}
\end{center}
\caption
{(Color online)
Diamond chain (a), frustrated ladder (b), square-{\kagome} (c), and checkerboard (d) lattices.
The circles indicate the lattice sites,
the thick (thin) lines indicate the hopping paths along the trap (connecting) bonds.
For the ideal diamond chain with $t_1=t_3$ we will use the notation
$t=t_1=t_3$, $t'=t_2$.}
\label{fig1}
\end{figure}

In what follows we focus for concreteness on some typical representatives
which were investigated in the context of strongly correlated systems earlier,
namely,
the diamond chain,\cite{diamond}
the frustrated ladder,\cite{bose_gayen,tanaka_idogaki}
the square-{\kagome},
and checkerboard lattices\cite{loc_mag1,loc_mag2}
(see Fig.~\ref{fig1}).
It is also convenient to denote sites by $m,p$,
where $m=1,\ldots,{\cal{N}}$ denotes unit cells
(${\cal{N}}=N/3,\,N/2,\,N/6,\,N/4$
for the diamond chain, frustrated ladder, square-{\kagome}, and checkerboard lattice, respectively)
and $p$ denotes sites inside the unit cell
(see Fig.~\ref{fig1}).
Note that for the lattices considered here
the number of unit cells is identical to the number of trapping cells.
However, the trapping cell may or may not be identical to the unit cell.

\section{Localized-electron states for $U=0$ and $U>0$}

\label{sec:loc}

Next we turn to the characterization of the complete manifold of highly degenerate ground states
for electron numbers $n=1,\ldots,{\cal{N}}\propto N$.
We start with the case $U=0$
considering the diamond chain
with $t=t_1=t_3$, $t'=t_2$ for concreteness.
Three one-electron bands,
\begin{eqnarray}
\varepsilon_{1}-\mu &=& -t^{\prime} \, , \nonumber \\
\varepsilon_{2,3}(\kappa)-\mu
&=& \frac{t^{\prime}}{2} \mp
 \sqrt{\left(\frac{t^{\prime}}{2}\right)^2+4t^2(1+\cos\kappa)} \, ,
\end{eqnarray}
are arranged as follows for $t^{\prime}>t^{\prime}_c=2t$:
$\varepsilon_{1}<\varepsilon_{2}(\kappa)<\varepsilon_{3}(\kappa)$.
Note that the lowest one-electron band $\varepsilon_{1}$ is completely
flat, i.e., $\kappa$-independent.
The corresponding states can be localized in real space.
We introduce the operators
$l_{m,\sigma}^{\dagger}=c_{m,2,\sigma}^{\dagger}-c_{m,3,\sigma}^{\dagger}$
[indices $2$ and $3$ denote the bottom and top sites on the vertical bond
(see Fig.~\ref{fig1}),
$m$ enumerates the unit cells]
which satisfy
$[H_{0\sigma},l^{\dagger}_{m,\sigma}]_{-}=\varepsilon_{1}\, l^{\dagger}_{m,\sigma}$.
Then all $2{\cal{N}}$ one-electron states belonging to the flat band
can be written as $l_{m,\sigma}^{\dagger}\vert 0\rangle$.
Application of $n$ distinct operators $l_{m,\sigma}^{\dagger}$ to $\vert 0\rangle$
yields $n$-electron states with energy $E_n = n\varepsilon_{1}$, $n=1,\ldots,{\cal{N}}$.
Note that all trapping cells are disconnected
and the degeneracy of these $n$-electron states is
\begin{equation}
g^{(0)}_{{\cal{N}}}(n)={2{\cal{N}} \choose n} \, .
\end{equation}
These arguments can be applied to other models:
for the energy of the flat band $\varepsilon_{1}-\mu$ we find
$-t^{\prime}$, $-2t^{\prime}$, $-2t^{\prime}$
for the frustrated ladder, square-{\kagome}, and checkerboard lattice, respectively.
We assume  $t^{\prime}/t> t'_c/t = 2,\,1,\,1$
for the frustrated ladder, square-{\kagome}, and checkerboard lattice, respectively.
For the square-{\kagome} and checkerboard lattice
an electron may be localized on smallest-area squares
and
$l_{m,\sigma}^{\dagger}
=c_{m,1,\sigma}^{\dagger}-c_{m,2,\sigma}^{\dagger}
+c_{m,3,\sigma}^{\dagger}-c_{m,4,\sigma}^{\dagger}$,
where the indices 1, \ldots, 4 denote the vertices of the square.

We now address the case $U>0$.
Since $H_U$ is a positive semidefinite operator for $U>0$,
it can only increase energies.
The states for which each trapping cell contains up to one electron
are exact eigenstates of the full Hamiltonian (\ref{01})
with the $U$-independent energy $E_n = n\varepsilon_{1}$
and thus they remain the ground states in the subspaces $n=2,\ldots,{\cal{N}}$
in the presence of a Hubbard repulsion $U>0$.
It is straightforward to count these localized $n$-electron
ground states and we find
\begin{equation}
g_{{\cal{N}}}(n)=2^{n} {{\cal{N}} \choose n} <g^{(0)}_{{\cal{N}}}(n) \, .
\label{gsDeg}
\end{equation}

The localized $n$-electron states are linearly independent
which can be proven using the arguments of Ref.~\onlinecite{schmidt}
(orthogonal class in the nomenclature of that reference).
These states are the only ground states in each subspace $n=1,\ldots,{\cal{N}}$.
This can be seen by recalling from spin systems
that a finite separation of the flat one-particle band from the next (dispersive) band
ensures completeness of the localized states.\cite{loc_mag2}
For the present models,
we can even control the energy gap by varying $t^{\prime}/t$,
in contrast to the sawtooth Hubbard chain.\cite{sawtooth}
The state with ${\cal{N}}$ electrons possesses perfect charge ordering
(each trapping cell is occupied by precisely one electron)
and therefore it can be understood as a particular realization of a Wigner crystal.\cite{Wu}
However, with respect to the spin orientations it has a huge degeneracy $2^{\cal{N}}$
(compare the degeneracy of the ground state, 
${\cal{N}}+1$,
at quarter filling when $n={\cal{N}}$ for the sawtooth chain\cite{flat_band_ferro,sawtooth}).

\section{Thermodynamics}

\label{sec:Thermo}

The localized-electron ground states in the sectors with $n \le {\cal{N}}$
have important implications for the low-temperature properties of the Hubbard model (\ref{01})
around a chemical potential $\mu_0=\mu-\varepsilon_{1}$:
Due to their huge degeneracy
they dominate the grand-canonical partition function at low temperatures.
Knowing the degeneracy (\ref{gsDeg})
of the ground states $g_{{\cal{N}}}(n)$
and their energy $E_n=n\varepsilon_{1}$
we can write the grand-canonical partition function as
\begin{equation}
\Xi(T,\mu,N)=\sum_{n=0}^{{\cal{N}}}g_{\cal{N}}(n)\, e^{-n\varepsilon_{1}/T}
=\left(1+2\, e^{-\varepsilon_{1}/T}\right)^{{\cal{N}}} \, .
\end{equation}
Note that only the combination $x=-\varepsilon_{1}/T=(\mu_0-\mu)/T$
enters the localized-electron contribution to any thermodynamic quantity.
The thermodynamic potential becomes
$\Omega(T,\mu,N)/{\cal{N}}=-T\ln(1+2\exp x)$
(the rhs.\ is valid for both finite $N$ and $N\to\infty$)
leading to simple expressions for thermodynamic quantities such as
\begin{eqnarray}
\frac{\nbar(T,\mu,N)}{{\cal{N}}}=\frac{2\, e^{x}}{1+2\, e^{x}}
\label{02}
\end{eqnarray}
for the average electron density
$\nbar(T,\mu,N)=\partial \Omega(T,\mu,N)/\partial \mu$,
\begin{eqnarray}
\frac{S(T,\mu,N)}{{\cal{N}}}=\ln(1+2\, e^{x})-\frac{2\,x\, e^{x}}{1+2\, e^{x}}
\label{03}
\end{eqnarray}
for the entropy
$S(T,\mu,N)=-\partial \Omega(T,\mu,N)/\partial T$,
or the specific heat in the grand-canonical ensemble
\begin{eqnarray}
\frac{C(T,\mu,N)}{{\cal{N}}}
&=&\frac{T\partial S(T,\mu,N)}{{\cal{N}}\partial T}
\nonumber\\
&=&\frac{2\,x^2\, e^{x}}{1+2\, e^{x}}
-\left(\frac{2\,x\, e^{x}}{1+2\, e^{x}}\right)^2.
\label{04}
\end{eqnarray}
The specific heat at constant $n$ vanishes identically,
$C(T,n,N)=0$, 
as can be verified explicitly using the
results (\ref{02}), (\ref{03}), and (\ref{04})
for $\nbar(T,\mu,N)$, $S(T,\mu,N)$, and $C(T,\mu,N)$.
Eqs. (5) -- (8) given above represent the contribution of the localized-electron states 
to the respective thermodynamic quantities. 
In section V we will demonstrate that these analytic expressions 
are an excellent description of the thermodynamics of the full model at low temperatures.

The average ground-state electron number $\nbar(0,\mu_0,N)$
exhibits a jump from ${\cal{N}}$ to zero as $\mu$ exceeds $\mu_0$.
Moreover,
at $\mu=\mu_0$ we have $x=0$
resulting in a finite residual entropy
$S(0,\mu_0,N)/{\cal{N}}=\ln 3\approx 1.0986$.
We note that thermodynamic quantities for all considered models
are identical up to a factor ${\cal{N}}/N$ and the concrete value  of
$\mu_0$, i.e., the low-temperature behavior is universal for the
whole family of lattices constructed by the rules formulated
in section \ref{sec:lat}.

\section{Numerical results}

Let us now present numerical results obtained by exact diagonalization
for finite lattices.
We set $t=1$ for convenience.
These calculations have been performed,
on the one hand,
to estimate the range of validity of the derived expressions for the 
low-temperature thermodynamics
and, on the other hand,
to study the states with $n > \cal N$ in more detail.
The numerical effort to diagonalize the
Hubbard model (\ref{01}) grows rapidly with $N$. It is therefore convenient
to consider also the limit $U \to \infty$ where doubly occupied sites can
be eliminated from the Hilbert space. Computations
were performed using different programs including J.~Schulenburg's
{\tt spinpack}\cite{spinpack} and a custom implementation of
the Householder algorithm.\cite{Householder}

Note first that the degeneracies of the localized ground states
calculated for various lattices and parameter sets
perfectly fit to the prediction $g_{{\cal{N}}}(n)$
of section~\ref{sec:loc}.
%
\begin{figure}
\begin{center}
\includegraphics[width=\columnwidth]{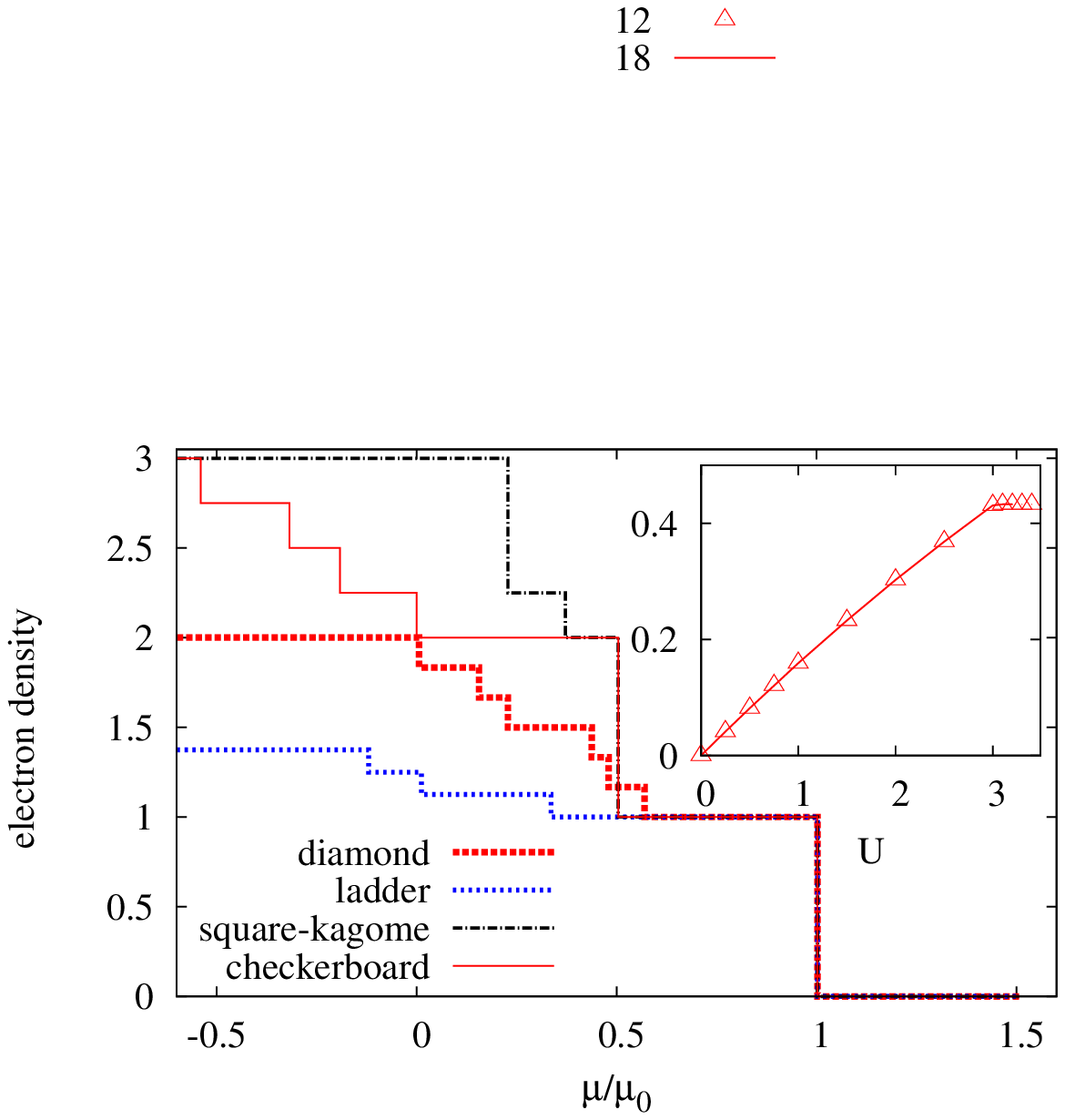}
\end{center}
\caption
{(Color online)
Electron density $\nbar/{\cal{N}}$ versus chemical potential $\mu/\mu_0$
for the diamond chain ($N=18$, $t^{\prime}=3$),
the frustrated ladder ($N=16$, $t^{\prime}=3$),
the square-{\kagome} lattice ($N=24$, $t^{\prime}=2$),
and
the checkerboard lattice ($N=16$, $t^{\prime}=2$)
for $U\to\infty$ and $T=0$.
The universal dependence (\ref{02}) at $T=0$ is given by $\theta(1-\mu/\mu_0)$.
Inset:
Charge gap $\Delta\mu/\mu_0$ at $\nbar/{\cal{N}}=1$ versus $U$
for the diamond chain [$N=12$ (triangles), $N=18$ (line), $t^{\prime}=3$].}
\label{fig2}
\end{figure}
The average electron density $\nbar/{\cal{N}}$ versus $\mu/\mu_0$ shown in Fig.~\ref{fig2}
exhibits a jump between 0 and 1 at $\mu/\mu_0=1$
and a plateau at $\nbar/{\cal{N}}=1$
with the width $\Delta\mu/\mu$.
The charge gap
$\Delta\mu=E({\cal{N}}+1)-2E({\cal{N}})+E({\cal{N}}-1)$
determines the region
in which the localized states exclusively control the ground-state behavior of the model (\ref{01}).
The inset in Fig.~\ref{fig2} shows that $\Delta\mu$ is almost size-independent.
The charge gap increases almost linearly with $U$ for small $U$, showing
that this is a correlation effect. For the diamond chain with
$t^{\prime}=3\,t$, the charge gap saturates at $\Delta\mu \approx 0.43\,t$
for $U \gtrsim 3\,t$.

\subsection{Thermodynamics for ideal geometry}

\begin{figure}
\begin{center}
\includegraphics[width=\columnwidth,angle=0]{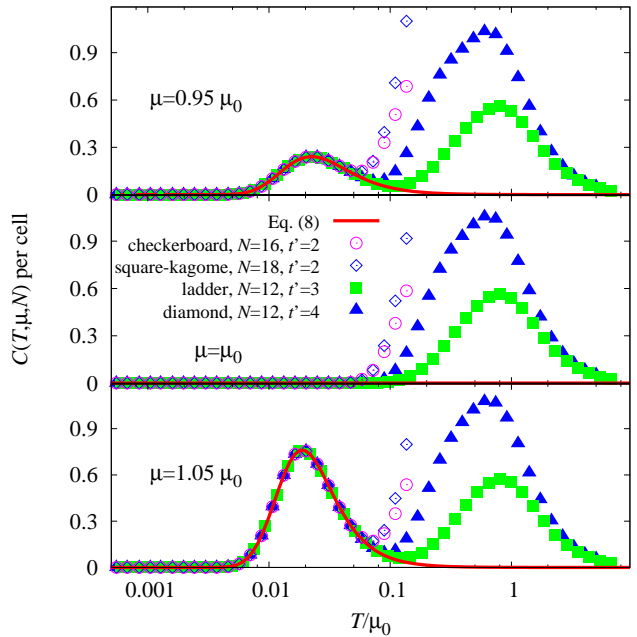}
\end{center}
\caption
{(Color online)
$C(T,\mu,N)/{\cal{N}}$ versus temperature $T/\mu_0$
in the limit $U\to\infty$
for the diamond chain 
(triangles),
the frustrated ladder 
(squares),
the square-{\kagome} lattice 
(diamonds),
and
the checkerboard lattice 
(circles).
We also show the universal dependence (\ref{04}) (lines).}
\label{fig3}
\end{figure}

Next we compare the specific heat $C(T,\mu,N)$
in the grand-canonical ensemble for
the diamond chain,
the frustrated ladder,
the square-{\kagome} lattice,
and
the checkerboard lattice
at $\mu = 0.95\,\mu_0$, $\mu_0$, and $1.05\,\mu_0$
in the limit $U \to \infty$
(Fig.~\ref{fig3}).
$C(T,\mu,N)/{\cal{N}}$ versus $T/\mu_0$ exhibits a universal 
additional low-temperature maximum
for the chemical potential around $\mu_0$.
This
low-temperature maximum emerges due to the manifold of localized-electron
states, whose energies are slightly split for
$\mu = 0.95\,\mu_0$ and $1.05\,\mu_0$ but still
well separated from energies of other higher energy states.
Indeed, the low-temperature maximum
is excellently described by the localized-electron formula (\ref{04})
(lines in Fig.~\ref{fig3}). 
Recall that the result (\ref{04}) is valid
for any $N$. Accordingly, there are no finite-size effects for the
low-temperature maximum of the specific heat. In Fig.~\ref{fig3},
one observes another maximum of $C(T,\mu,N)$ at higher temperatures.
This maximum collects all states which are not localized-electron states
and thus depends on details of the model.
If $\mu\to\mu_0$ the low-temperature maximum shifts to lower $T$
and is even much better separated from the high-temperature maximum.
At $\mu=\mu_0$, we have $x=0$. Consequently,
the contribution (\ref{04}) of the localized-electron states
to the specific heat $C$ vanishes identically.
Therefore, the
temperature where the numerical data for $C(T,\mu_0,N)$
begin to deviate from zero (see the middle panel of Fig.~\ref{fig3})
represents a characteristic temperature
below which the localized states exclusively control the thermodynamic
behavior of the model (\ref{01}).

\begin{figure}
\begin{center}
\includegraphics[width=\columnwidth,angle=0]{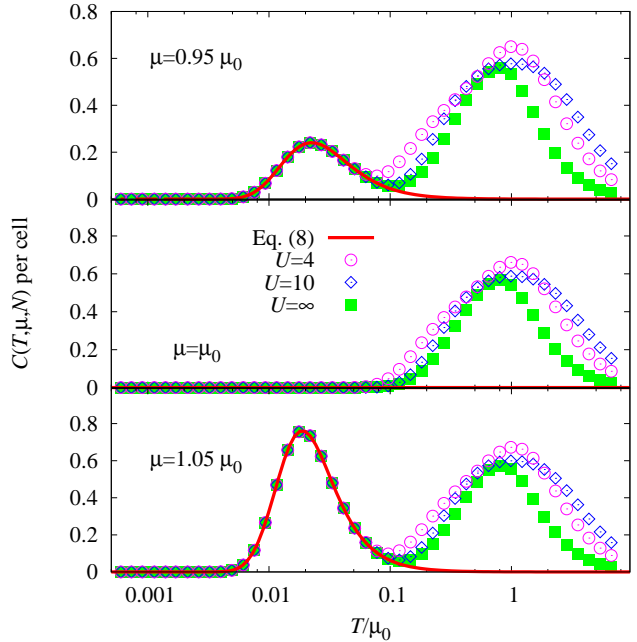}
\end{center}
\caption
{(Color online)
$C(T,\mu,N)/{\cal{N}}$ 
for the frustrated ladder with $N=12$ and $t^{\prime}=3$
for different values of $U$.
We also show the universal dependence (\ref{04}) (lines).}
\label{fig4}
\end{figure}

Fig.~\ref{fig4} illustrates the influence of different values of $U$,
using the example of the frustrated ladder with $N=12$ and $t^{\prime}=3$.
The low-temperature maximum in $C(T,\mu_0,N)$ is independent of $U$,
as expected. By contrast, the high-temperature maximum does depend on $U$
and it extends to lower temperatures for smaller values of $U$. Nevertheless,
the localized-electron states are still well separated from
the other states
for the values of $U \ge 4$ and $\mu=0.95\,\mu_0$, $1.05\,\mu_0$
shown in Fig.~\ref{fig4}.

\begin{figure}
\begin{center}
\includegraphics[width=\columnwidth,angle=0]{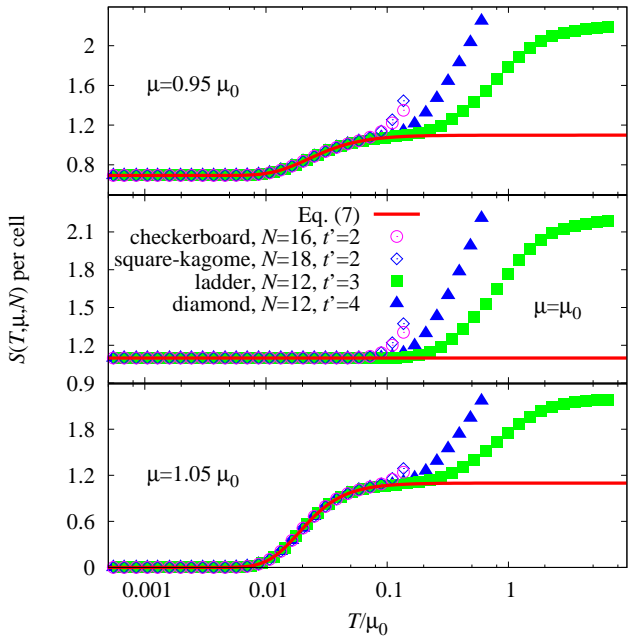}
\end{center}
\caption
{(Color online)
Entropy $S(T,\mu,N)/{\cal{N}}$ versus temperature $T/\mu_0$
in the limit $U\to\infty$
for the diamond chain 
(triangles),
the frustrated ladder 
(squares),
the square-{\kagome} lattice 
(diamonds),
and
the checkerboard lattice 
(circles).
We also show the universal dependence (\ref{03}) (lines).}
\label{fig5}
\end{figure}

Fig.\ \ref{fig5} shows the entropy $S$ for the same lattices and parameters
as in Fig.\ \ref{fig3}. Since $C(T,\mu,N)$ is a temperature derivative
of $S(T,\mu,N)$, the features in the latter quantity correspond to those
in the former. 
In particular a steep slope in $S(T,\mu,N)$
corresponds to a maximum of $C(T,\mu,N)$.
However, there is one additional piece of information
in $S(T,\mu,N)$, namely its value in the low-temperature limit:
for $\mu = \mu_0$, $S(T,\mu,N)/{\cal N}$ tends to $\ln 3\approx 1.0986$ for $T \to 0$
(see section \ref{sec:Thermo}). 
If the chemical potential $\mu$ is in the charge-gap region, 
i.e., for $\mu_0-\Delta \mu < \mu < \mu_0$, 
the entropy per unit cell stays also finite for $T\to0$
(compare the upper panel of Fig.\ \ref{fig5}).
For these values of $\mu$  the ground state 
is the charge-ordered state (Wigner crystal)
with $n={\cal N}$ electrons and accordingly
exhibits  a ground-state degeneracy of $g_{\cal N}({\cal N}) = 2^{\cal N}$ 
due to the independence of the spin orientations of the ${\cal{N}}$ electrons, 
see Eq.~(\ref{gsDeg}).
Indeed, Eq.~(\ref{03}) shows that the entropy per cell
$S/{\cal{N}}\to\ln 2\approx 0.6931$ when $x\to\infty$, i.e.,
$T\to0$ and $\mu < \mu_0$.

\subsection{Deviations from ideal geometry}

\begin{figure}
\begin{center}
\includegraphics[width=\columnwidth,angle=0]{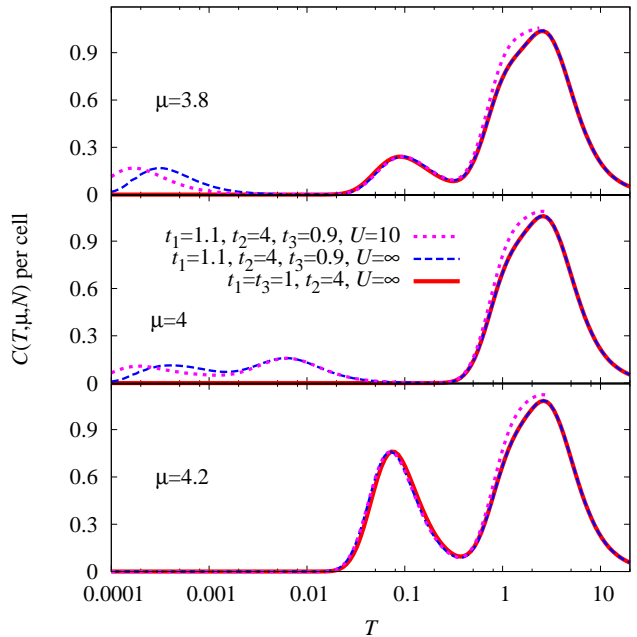}
\end{center}
\caption
{(Color online)
$C(T,\mu,N)/{\cal{N}}$ for the ideal diamond chain
($t^{\prime}=4$, $t=1$) and the
distorted diamond chain ($t_1=1.1$, $t_2=4$, $t_3=0.9$).
The system size is $N=12$ in all cases.}
\label{fig6}
\end{figure}

In real compounds
the conditions under which the localized states exist may be not strictly fulfilled.
It is therefore important to study deviations from `ideal geometry'.
Such deviations or distortions in general lift the
ground-state degeneracy.
Nevertheless, one may expect that localized-state effects survive in
a certain temperature range if the distortion is sufficiently small such that the
originally degenerate energy levels remain close to each other.
We investigate this issue in more detail using the example of the
distorted diamond chain ($N=12$, $t^{\prime}=4$) 
where we distinguish the hopping terms
running from north-west to south-east $t_1>t$ ($t_1=1.1$)
and those 
running from south-west to north-east $t_3<t$ ($t_3=0.9$),
see Fig.~\ref{fig1}. This generalization
is inspired by the set of exchange
interactions proposed for azurite in Ref.~\onlinecite{azurite}.
Fig.~\ref{fig6} shows numerical results for $N=12$, $U=\infty$ and $10$
in comparison to the data for the undistorted case 
which are identical to those 
shown in Fig.~\ref{fig3}. The effect of the distortion is evidently
very small for $T \gtrsim 0.04$.
In particular, the additional low-temperature maximum in the grand-canonical
specific heat $C(T,\mu,N)/{\cal{N}}$ for $\mu=3.8$ and $4.2$
(corresponding to $\mu=0.95\,\mu_0$ and $1.05\,\mu_0$ in the
undistorted case) is essentially unaffected. This implies that
an important fingerprint of the highly degenerate localized states
survives a small distortion.
Additional features emerge in Fig.~\ref{fig6} for the distorted diamond
chain in the region $T \lesssim 0.01$ at $\mu=3.8$ and
for $T \lesssim 0.1$
at $\mu=4$. It is evident from Fig.~\ref{fig6} that
these features depend on the value of $U$. Comparison with smaller
system sizes for $U=10$ (not shown) exhibits also finite-size effects for
$T \lesssim 0.001$ at $\mu=3.8$ and
for $T \lesssim 0.01$ at $\mu=4$.
One may therefore speculate that the additional low-temperature
maximum observed in $C(T,\mu,N)$ for $\mu=4$ around $T \approx 0.006$
in Fig.~\ref{fig6} survives the thermodynamic limit, but it is difficult
to infer the behavior in the thermodynamic limit at even lower temperatures.

The specific heat $C(T,n,N)$ in the {\em canonical} ensemble, i.e., for a fixed
number of electrons $n$, may be particularly
relevant from the experimental point of view.
In section \ref{sec:Thermo}
we have already used the results for the grand-canonical
ensemble to point out
that the contribution of the localized-electron states to
$C(T,n,N)$ always vanishes identically in the ideal situation.
This is also evident if one considers a fixed number of electrons:
in this case all localized $n$-electron states have the same energy
$E$ and using the representation
$C=\left(\langle E^2 \rangle - \langle E \rangle^2\right)/T^2$
in terms of the fluctuations of $E$ one finds $C(T,n,N)=0$.

\begin{figure}
\begin{center}
\includegraphics[width=\columnwidth,angle=0]{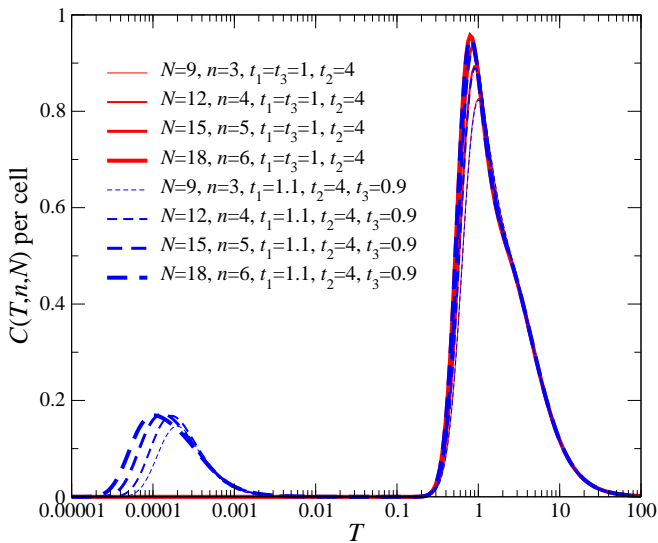}
\end{center}
\caption
{(Color online)
Specific heat
$C(T,n,N)/{\cal{N}}$ for the ideal diamond chain
($t^{\prime}=4$, $t=1$) and the
distorted diamond chain ($t_1=1.1$, $t_2=4$, $t_3=0.9$)
with $n=N/3$ electrons for $N=9,\;12,\;15,\;18$.
The Hubbard repulsion is $U=10$.}
\label{fig7}
\end{figure}

We use again the example of the (distorted) diamond chain to take
a closer look at the behavior of $C(T,n,N)$. The case of
the Wigner crystal, i.e., $n={\cal N}$ may be particularly interesting.
Fig.~\ref{fig7} shows numerical results for the
specific heat $C(T,n,N)/{\cal{N}}$ of both the ideal and
the distorted diamond chain with $n={\cal N}=N/3$ electrons.
First, we observe that $C(T,n,N)$ of the ideal diamond chain
(solid lines in Fig.~\ref{fig7})
is indeed indistinguishable from zero for $T\lesssim 0.1$
and accordingly there is no low-temperature maximum. 
By contrast, for the distorted diamond chain the
exact degeneracy of the ground-state manifold is removed,
leading to the recovery of a low-temperature maximum.
For the parameters used in Fig.~\ref{fig7}, this low-temperature
maximum appears at $T \approx 10^{-4}$ (see dashed lines
in Fig.~\ref{fig7}), reflecting the fact
that the detuning $t_3=0.9 \ne t_1=1.1$ leads only
to a very small splitting of the originally $2^n$-fold degenerate
ground-state manifold. There is another high-temperature
maximum at temperatures of order one
in Fig.~\ref{fig7} where the results for the distorted diamond chain
are indistinguishable from those for the ideal situation.
The double-peak structure found in the canonical specific heat
$C(T,n,N)$ of the distorted situation is qualitatively remarkably similar to
the double-peak structure in the grand-canonical specific heat
$C(T,\mu,N)$ observed e.g.\ for $\mu=3.8$ in Fig.~\ref{fig6}.

While finite-size effects are mostly unimportant in the grand-canonical
ensemble, they are clearly relevant in the low-temperature
region of Fig.~\ref{fig7} and visible even in the region of the
high-temperature maximum. Such enhanced finite-size effects are
typical for computations performed in the canonical ensemble where
one considers only a fixed number of electrons $n$.

\section{Magnetic properties}

We return now to the ideal geometry and discuss 
ground-state magnetism.
Since the trapping cells are independent of each other
(i.e., they do not have common sites),
the cells can be occupied independently by electrons with spin up or spin
down for $n=1,\ldots,{\cal{N}}$.
As a result, for $n \le {\cal{N}}$ 
the majority of the degenerate ground
states 
of the Hamiltonian (\ref{01}) on the
considered lattices is nonmagnetic.
Indeed,
using the localized-electron picture of section \ref{sec:loc},
it is straightforward to show for the total spin operator 
$\bm{S}=(S^x,S^y,S^z)$
that
\begin{equation}
\frac{\langle \bm{S}^2\rangle_n}{N^2}
=\frac{3\langle (S^z)^2\rangle_n}{N^2}
= \frac{3n}{4\,N^2}
\stackrel{N \to \infty}{\longrightarrow} 0
\end{equation}
for $n \le {\cal{N}}$,
where the limit $N \to \infty$ is taken for a fixed electron density $n/N$.
Here $\langle \ldots\rangle_n$ denotes the average over
all ground states
in the sector with $n$ electrons.

The situation changes fundamentally at $n={\cal{N}}+1$.
If $U$ is small
(with respect to the energy gap between the flat band and the next band;
this gap can be controlled by the ratio $t^{\prime}/t$)
an extra electron may lead to formation of a complicated many-body state
with an energy that increases with growing $U$.
However, if $U$ is large enough, $U>U_c({\cal{N}}+1)$,
it might be energetically favorable to avoid double occupancy
by putting one extra electron in the lowest-energy state
of the next  (dispersive) one-electron band
in addition to the $\cal N$ localized electrons.
The magnitude of $U_c$ depends on the system under consideration. 
It is possible to find an explicit expression for such an eigenstate:
\begin{equation}
\vert\varphi_{{\cal {N}}+1}\rangle
\propto
\beta_{\kappa_0,\uparrow}^{\dagger}
l_{{\cal{N}},\uparrow}^{\dagger}\ldots l_{{1},\uparrow}^{\dagger}\vert 0\rangle
\, .
\label{eq:10}
\end{equation}
Here $\beta_{\kappa_0,\sigma}^{\dagger}$ creates an 
electron in the lowest dispersive band 
with momentum $\kappa_0$ and spin $\sigma$. 
The state $\vert\varphi_{{\cal {N}}+1}\rangle$ is fully polarized,
i.e., 
\begin{equation}
\langle\varphi_{{\cal{N}}+1}\vert \bm{S}^2\vert\varphi_{{\cal{N}}+1}\rangle
=
\frac{{\cal{N}}+1}{2}\,\left(\frac{{\cal{N}}+1}{2}+1\right) \, ,
\end{equation}
and it has a $U$-independent energy
${\cal{N}}\varepsilon_1+\varepsilon_2(\kappa_0)$.
Other states belonging to a spin-$\frac{{\cal{N}}+1}{2}\;$ SU(2)-multiplet
can be obtained by applying
$S^{-}=\sum_{i}c_{i,\downarrow}^{\dagger}c_{i,\uparrow}$
to this state (Kramers degeneracy).
This multiplet can also be obtained by applying 
$S^{+}=\sum_{i}c_{i,\uparrow}^{\dagger}c_{i,\downarrow}$ to
the spin down counterpart of Eq.~(\ref{eq:10}).

\begin{table}
\begin{center}
\caption
{Number of electrons $n>1$ for which there is a unique saturated
  ferromagnetic ground-state
  multiplet in the limit $U\to\infty$ on the specified
  lattice subject to periodic boundary conditions. Note that saturated
  ground-state ferromagnetism for the considered lattices of $N$ sites
  occurs only for a sufficiently large number of electrons ${\cal{N}}<n<N$
  (apart from the trivial case $n = 1$). Note that for the
square-{\kagome} lattice  with $N=24$ sites the ground states for
$n=16,17,18,19$
have not been calculated, because
of the large size of the
Hamiltonian matrix.
\label{table}}
\vspace{5mm}
\begin{tabular}{|c||c|} \hline
lattice                                  & number of electrons         \\ \hline \hline 
diamond,          $N=12$ (${\cal{N}}=4$) & $n=5,7,8,9,11$              \\ \hline
diamond,          $N=18$ (${\cal{N}}=6$) & $n=7,9,11,12,13,15,17$      \\ \hline
ladder,           $N=12$ (${\cal{N}}=6$) & $n=7,9,11$                  \\ \hline
ladder,           $N=16$ (${\cal{N}}=8$) & $n=9,11,13,15$              \\ \hline
square-{\kagome}, $N=18$ (${\cal{N}}=3$) & $n=4,5,7,8,15,17$           \\ \hline
square-{\kagome}, $N=24$ (${\cal{N}}=4$) & $n=7,8,9,\ldots,20,21,23$   \\ \hline
checkerboard,     $N=16$ (${\cal{N}}=4$) & $n=7,8,9,15$                \\ \hline
\end{tabular}
\end{center}
\end{table}

Numerically, we find such ferromagnetic ground states for large $U$ also
for bigger ${\cal N} < n < N$.
For finite systems the numbers of electrons $n$ 
for which ferromagnetic ground states appear 
depend on the lattice size and the boundary conditions.\cite{watanabe,spiral}
For all lattices considered in this paper and the imposed periodic boundary conditions 
we find fully polarized ground states for particular values of the electron number $n>{\cal{N}}$
for sufficiently large $U > U_c$.
In Table~\ref{table}
we list some combinations of finite lattices  and numbers of electrons $n$
for which the ground state is fully polarized for $U\to\infty$.
Note that the state (\ref{eq:10}) is incompatible with the
boundary conditions of the $N=24$ square-{\kagome} and the $N=16$
checkerboard lattice such that there is no saturated ground-state
ferromagnetism for $n=5$ on these two lattices.
In addition, 
we give as an example the strength of correlation $U> U_c$ 
that is needed to realize ground-state ferromagnetism 
for $n={\cal{N}}+1$ for the frustrated ladder with $N=16$ sites: 
we found  $U_c \approx 0.42$ and $12.78$ for $t^{\prime}=2.1$ and $4$, 
respectively.

This kind of ferromagnetism that occurs only for sufficiently strong on-site repulsion
(note that it is not flat-band ferromagnetism that occurs for any $U>0$)
was discussed also earlier for some one-dimensional systems.\cite{watanabe,tanaka_idogaki,spiral}
Moreover we mention that the saturated ferromagnetism found 
for the largest electron number $n=N-1$ 
corresponds to Nagaoka's well-known
result.\cite{flat_band_ferro,nagaoka}
However,
we emphasize that the occurrence of ferromagnetism 
for other electron numbers in the range ${\cal N} < n < N$
is a generic feature of the considered lattices
and is therefore not restricted to one-dimensional systems.

\section{Conclusions}

In summary,
we have given an exact solution for the ground-state properties
of a correlated many-electron system on a class of lattices
in a certain range of the chemical potential $\mu$.
We have studied the low-temperature thermodynamics
which for $\mu$ around $\mu_0$ is controlled just by the manifold 
of localized ground states.
In particular, we have presented
explicit expressions for the low-temperature behavior of the  
grand-canonical partition function and related quantities such as the
grand-canonical specific heat $C(T,\mu,N)$.
The localized-electron features have no finite-size effects 
and are universal, i.e., they are the same for
the whole class of lattices.

Apart from
the high-temperature maximum that 
is typical for systems with a bounded energy
spectrum we find an additional low-temperature maximum in 
$C(T,\mu,N)$ if the chemical potential $\mu$ deviates
slightly
from $\mu_0$.
This extra maximum survives
under small deviations from ideal lattice geometry. Moreover,
for the canonical specific heat $C(T,n,N)$
the lifting of degeneracy caused by deviations from ideal lattice geometry actually
{\em gives rise} to an additional
low-temperature maximum in the specific heat. Thus,
any splitting of the highly degenerate localized electron states, be it by a
deviation of the 
chemical potential from $\mu_0$ 
or by a deviation from ideal lattice geometry, 
leads to an extra low-temperature maximum
in the specific heat as a characteristic fingerprint 
of the localized-electron states.

Furthermore, we argued that in contrast to the flat-band ferromagnets\cite{flat_band_ferro}
there is no ground-state ferromagnetism in the present class of models
if the electron density satisfies $n/{\cal N} \le 1$.
%
For electron numbers ${\cal{N}} < n < N$
we observe saturated ground-state
ferromagnetism for all considered lattices,
including square-{\kagome} and checkerboard lattices. An explicit
expression for the saturated ferromagnetic ground state in the
sector $n={\cal N}+1$ is given in Eq.~(\ref{eq:10}).

Finally we note that the $t-J$ model on the considered lattices
may exhibit similar localized ground states in the subspaces with
$n=1,\ldots,{\cal{N}}$ electrons for values of
the spin exchange interaction $J$
up to about the hopping integral $t$.
Another interesting variation on the considered models refers to multiorbital systems.\cite{Wu}
For instance,
the one-orbital Hubbard model on a frustrated ladder
can be related to a two-orbital Hubbard model on a simple chain
with a hybridization term corresponding to the hopping on a rung.
This relation between one-orbital models and multiorbital models
extends the region of applicability of the localized-states scenario.

\section*{Acknowledgments}

We thank J.~J\c{e}drzejewski for discussions.
O.D.\ acknowledges the financial support of the DAAD.
Financial support of the DFG is gratefully acknowledged
(project RI 615/18-1 and a Heisenberg fellowship
for A.H.\ under project HO~2325/4-1).


\begin{thebibliography}{999}
\bibitem{extension}
More general trapping cells
such as non-bipartite cells or cells with non-equivalent sites and bonds
are possible as well,
cf.\ Refs.~\onlinecite{batista,sawtooth}.
\end{thebibliography}
\end{document}